\documentclass[showpacs,showkeys,aps,prb,twocolumn]{revtex4}
\usepackage[english]{babel}
 \usepackage{epsfig}
\usepackage{graphicx} \usepackage{amsmath}
\usepackage{amssymb}
\usepackage{subfigure}
\usepackage{bm}
\usepackage{subfigure}

\begin{document}

\title{Doped orbitally-ordered systems: another case of phase
separation}

\author{ K.~I.~Kugel~\cite{affUK}}

\affiliation{Institute for Theoretical and Applied
Electrodynamics, Russian Academy of Sciences, Izhorskaya Str. 13,
Moscow, 125412 Russia}

\author{A.~L.~Rakhmanov~\cite{affUK}}

\affiliation{Institute for Theoretical and Applied
Electrodynamics, Russian Academy of Sciences, Izhorskaya Str. 13,
Moscow, 125412 Russia}

\author{A.~O.~Sboychakov}

\affiliation{Institute for Theoretical and Applied
Electrodynamics, Russian Academy of Sciences, Izhorskaya Str. 13,
Moscow, 125412 Russia}

\author{D.~I.~Khomskii~\cite{affUK}}

\affiliation{$II.$ Physikalisches Institut, Universit\"at zu
K\"oln, Z\"ulpicher Str. 77, 50937 K\"oln, Germany}

\begin{abstract} A possible mechanism of electronic phase
separation in the systems with orbital ordering is analyzed. We
suggest a simple model taking into account an interplay between
the delocalization of charge carriers introduced by doping and the
cooperative ordering of local lattice distortions. The proposed
mechanism is quite similar to the double exchange usually invoked
for interpretation of phase separation in doped magnetic oxides
like manganites, but can be efficient even in the absence of any
magnetic ordering. It is demonstrated that the delocalized charge
carriers  favor the formation of nanoscale inhomogeneities with
the orbital structure different from that in the undoped material.
The directional character of orbitals leads to inhomogeneities of
different shapes and sizes.

\end{abstract}



\date{\today}

\maketitle

\section{Introduction}\label{Intr}

The existence of superstructures is a characteristic feature of
magnetic oxides, in particular those containing ions with orbital
degeneracy, i.e., Jahn-Teller (JT) ions. In the crystal lattice,
the JT ions usually give rise to the orbital ordering
(OO)~\cite{KK,KaVe}. The OO is typical of insulating compounds.
The electron or hole doping can destroy OO since the itinerant
charge carriers favor the formation of a metallic state without
OO. However, at low doping level, we have a competition between
the charge localization and metallicity. It is well known that
such a competition can lead to the so-called electronic phase
separation (PS) with nanoscale
inhomogeneities~\cite{dagbook,Nag,Kak}. This phenomenon is often
observed, e.g., in doped manganites and is usually related to some
specific type of magnetic ordering, antiferromagnetic insulator
versus ferromagnetic metal. In the usual treatment of PS, the OO
is not taken into account (see, however, the discussion concerning
isolated orbital and magnetic
polarons~\cite{KilKha,MiKhoSaw,KhaOk,vdBKhaKho}). Here we study
the effect of OO on PS employing minimal models including
itinerant charge carriers at the OO background, and show that at
small doping the PS may appear in systems with orbital degeneracy
even without taking into account magnetic structure. We consider
this effect using two versions of the models. First, in Section II
we study a symmetrical model analogous to the Kondo-lattice model
in the double exchange limit, where the orbital variables play a
role of local spins. Namely, it is supposed that localized
electrons create lattice distortions, leading to the formation of
OO. The conduction electrons or holes, introduced by doping, move
on OO background. In the second version (Section III), we take
into account the specific symmetry of $e_g$ type for doped
electrons. For both versions, we demonstrate the possible
instability of a homogeneous ground state against the formation of
inhomogeneities. As a result, additional charge carriers
introduced by doping favor the formation of nanoscale
inhomogeneities with the orbital structure different from that in
the undoped material. In Section IV, we determine the shapes and
sizes of such inhomogeneities and demonstrate that depending on
the ratio of the electron hopping integral $t$ and the
interorbital coupling energy $J$, the shape can vary drastically.
For the two-dimensional case, in particular, there exists a
critical value of $t/J$, corresponding to the abrupt transition
from nearly circular to needle-like inhomogeneities. This is a
specific feature of orbital case: the directional character of
orbitals brings about the unusual and very rich characteristics of
inhomogeneous states.

\section{Symmetrical model}\label{SymModel}

Let us consider the system with JT ions having double-degenerate
state. This degeneracy can be lifted by local lattice distortions,
giving rise to two different ground states of each ion, $a$ or $b$
(e.g. $a$ ($b$) state corresponds to elongation (compression) of
anion octahedra). The states $a$ and $b$ of the ion $\mathbf{n}$
determine the corresponding  orbital states of a charge carrier at
this ion. In general case, each ion can be characterized by a
linear combination of basis $a$ and $b$ states, described by an
angle $\theta$ \begin{equation}\label{theta}
|\theta\rangle=\cos\frac{\theta}{2}|a\rangle+\sin\frac{\theta}{2}|b\rangle\,.
\end{equation} The local distortions can interact with each other
leading to some regular structure. In the simplest symmetrical
case, the interaction Hamiltonian can be written in a
Heisenberg-like form \begin{equation}\label{H_Heis}
H_{\text{OO}}=J\sum_{\langle{\bf n}{\bf m}\rangle}\bm{\tau}_{\bf
n}\bm{\tau}_{\bf m}\,, \end{equation} where
$\bm{\tau_n}=\{\tau^x_{\mathbf{n}}$, $\tau^z_{\mathbf{n}}\}$ are
the Pauli matrices, and $a$ and $b$ states of the ion ${\bf n}$
correspond to eigenvectors of operators $\tau^z_{\bf n}$, with
eigenvalues $1$ and $-1$, respectively. For
Hamiltonian~\eqref{H_Heis}, two simplest kinds of ordering are
possible: ferro-OO (the same state at each site) and antiferro-OO
(alternating states at neighboring sites). In the absence of the
charge carriers, the ground state is antiferro-OO if $J>0$ and
ferro-OO if $J<0$. Of course, in real materials with Jahn-Teller
ions, the orbital Hamiltonians are more complicated, but the
analysis based on the model~\eqref{H_Heis} seems to be sufficient
to reproduce the essential physics related to the orbital
ordering.

Under doping, itinerant charge carriers appear in the system, so
the density of charge carries $n\neq 0$. We assume that the charge
carriers are doped into double-degenerate states and move on the
OO background determined by localized electrons. (As we argue
below, the main results will be also applicable to the case where
the same electrons, e.g. $e_g$ electrons, are responsible both for
the OO and for conduction due to doping into these $e_g$ states).
The values of electron hopping integrals should depend on the
states of the neighboring lattice sites. The electron Hamiltonian
can be written as \begin{equation}\label{H_el}
H_{\text{el}}=-\sum_{\langle{\bf n}{\bf
m}\rangle,\alpha,\beta,\sigma}t^{\alpha\beta}\left(P_{{\bf
n}\alpha\sigma}^{\dag}a^{\dag}_{{\bf n}\alpha\sigma}a_{{\bf
m}\beta \sigma}P_{{\bf m}\beta\sigma}+h.c.\right), \end{equation}
where, $a^{\dag}_{{\bf n}\alpha\sigma}$, $a_{{\bf n}\alpha\sigma}$
are creation and annihilation operators for the charge carriers at
site ${\bf n}$ with spin projection $\sigma$ at orbital $\alpha$.
Having in mind that we are dealing with a strongly correlated
electron system, we introduced in \eqref{H_el} projection
operators $P$ excluding a double occupation of lattice sites (we
consider the case $n<1$). Below, analyzing the electron
contribution to the total energy, we shall consider square lattice
in the two-dimensional (2D) case and cubic lattice in the
three-dimensional (3D) case using the tight-binding approximation.
For the spectrum of charge carriers, we have
\begin{equation}\label{E(k)} E^{\alpha\beta}({\bf
k})=-t^{\alpha\beta}\frac{1}{D}\sum_{i=1}^D\cos k_i
=t^{\alpha\beta}\xi({\bf k})\,, \end{equation} where $D$ is the
space dimensionality and $k_i$ are the components of wave vector
${\bf k}$ in the units of inverse lattice constant $1/d$.

In our model, doped electrons at a JT distorted site $a$ or $b$
are in the corresponding orbital state $|a\rangle$ or $|b\rangle$.
We can introduce three hopping integrals: $t^{aa}$, $t^{bb}$, and
$t^{ab}=t^{ba}=t'$. For simplicity, let us assume that
$t^{aa}=t^{bb}=t$. Then, we have a competition of two factors: the
formation either of a wider electron band or of an optimum OO
type.

At the site $\mathbf{n}$ in the state $\theta$, the charge carrier
has an orbital state $|\theta>$, described by Eq.~\eqref{theta}.
The hopping integral between the sites characterized by orbital
states $|\theta_1\rangle$ and $|\theta_2\rangle$ can be written as
\begin{equation}\label{t_theta}
t^{\theta_1\theta_2}=t\cos{\frac{\theta_1 -\theta_2}{2}}
+t'\sin{\frac{\theta_1+\theta_2}{2}}
\end{equation}
First, we
consider a homogeneous state, assuming that the orbital structure
corresponds to the alternation of $|\theta_1\rangle$ and
$|\theta_2\rangle$ orbitals. In the mean-field approximation, we
can represent the total energy per site as
\begin{equation}\label{Etot1}
E_{tot}(\theta_1,\theta_2)=zt^{\theta_1\theta_2}\varepsilon_0(n)
+\frac{zJ}{2}\cos(\theta_1-\theta_2)\,,\;\;\varepsilon_0(n)<0\,,
\end{equation}
where $z$ is the number of nearest neighbors, and
dimensionless kinetic energy $\varepsilon_0(n)$ is determined by
the type of the crystal lattice. A specific form of
$\varepsilon_0(n)$ for different cases will be discussed below. We
assume in this Section that $t$, $t'$, and, therefore,
$t^{\theta_1\theta_2}$, do not depend on the direction of hopping.
In this isotropic case, $\varepsilon_0(n)$ does not depend on
$\theta_{1,2}$, and we can easily calculate the orbital structure
by minimization of total energy~\eqref{Etot1} with respect to
angles $\theta_1$ and $\theta_2$. At relatively large doping, when
$t|\varepsilon_0(n)|>2J$, we have ferro-OO state with
$\theta_1=\theta_2=\pi/2$. In  the opposite case,
$t|\varepsilon_0(n)|<2J$, the minimization yields:
$\theta_2=\pi-\theta_1$, and \begin{equation}\label{thetaMin}
\theta_1=\arcsin\left(\frac{t|\varepsilon_0(n)|}{2J}\right)\,,
\;\;\frac{t|\varepsilon_0(n)|}{2J}<1\,. \end{equation} The total
energy of such a canted orbitally ordered state is
\begin{equation}\label{Etot2}
E_{tot}=zt'\varepsilon_0(n)-\frac{zt^2}{4J}
\varepsilon_0^2(n)-\frac{zJ}{2}\,. \end{equation} Note that if
$\varepsilon_0(n)=nf(n)$, where $f(n)$ varies slowly with $n$,
then $E_{tot}$ can have a negative curvature, at least at small
$n$, which is a signature of an instability of a homogeneous
orbitally ordered state (negative compressibility).

Let us now determine function $\varepsilon_0(n)$ and analyze the
dependence of the total energy on doping. For the tight-binding
spectrum~\eqref{E(k)} of electrons in the lattice of the
dimension $D$, the density of states $\rho_0(E)$, has the form
\begin{equation}\label{rho0}
\rho_0(E)=\!\int\!\frac{d\mathbf{k}}{(2\pi)^D}\delta(E-\xi(\mathbf{k}))=%
\!\int\limits_0^{\infty}\!\frac{ds}{\pi}\cos(Es)J_0^D\!\left(\frac{s}{D}\right)\!,
\end{equation} where $J_0$ is the Bessel function. Then we have

\begin{equation}\label{epsilon}
\varepsilon_0(n)=\int_{-1}^{\mu(n)}\!\!\!\!\!dE\,E\rho_0(E)\,,
\end{equation} with the chemical potential $\mu$ given by equation
$n=\int_{-1}^{\mu}dE\rho_0(E)$.

At small doping, $n\ll1$, it is possible to write
$\varepsilon_0(n)$ in a simple explicit form. In 2D case,
$\varepsilon_0(n)\approx-n+\pi n^2/2$, and the total energy then
reads
\begin{equation}\label{Etot2Da}
E_{tot}\approx-zt'n-z\left(\frac{t^2}{4J}-\frac{\pi
t'}{2}\right)n^2-\frac{zJ}{2}\,.
\end{equation}

From Eq.~\eqref{Etot2Da}, we find that $d^2E_{tot}/dn^2 < 0$ if
\begin{equation}
\frac{t}{J}>\frac{2\pi t'}{t},
\end{equation}
This implies an instability of the homogeneous orbitally canted
state toward the phase separation into phases with ferro- and
antiferro-orbital ordering. The situation  here is quite similar
to that for the usual double exchange~\cite{KaKhoMo}, which
corresponds to the case $t'=0$. At relatively large $t'$, when
$2\pi t'/t>t/J$, a homogeneous state is stable in the whole range
of doping.

\begin{figure} \begin{center}
\includegraphics*[width=0.45\textwidth]{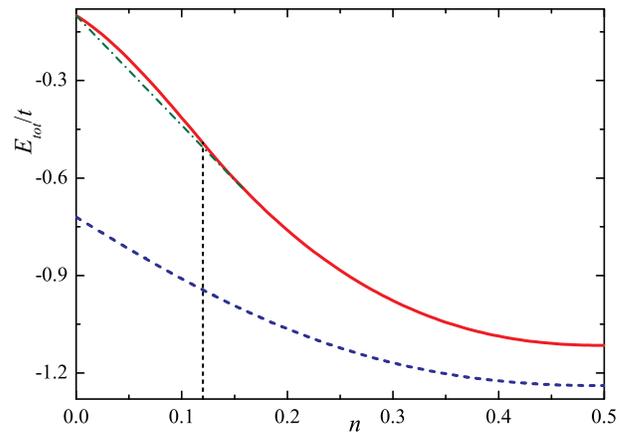} \end{center}
\caption{\label{FigEtot2D} (Color online) Two types of behavior of
the energy of homogeneous state ~\eqref{Etot2} in 2D as function
of doping $n$: with a region of negative curvature (red solid
line, $J/t=0.05$), and without it (blue dashed line, $J/t=0.35$);
$t'=0.5t$ for both curves. For the values of parameters
corresponding to the solid curve, the orbitally canted state
existing on the left-hand side of vertical line is  unstable
toward a phase separation. The Maxwell construction in a region of
phase separation is shown by green dot-dashed line. The
homogeneous state corresponding to the blue dashed curve is stable
in the whole range of doping.} \end{figure}

Taking $\varepsilon_0(n)\approx-n$ in Eq.~\eqref{thetaMin} we get
a rough estimate for a region of phase separation:
\begin{equation}\label{une1}
0<n\lesssim\frac{2J}{t}.
\end{equation}
So, the orbitally canted state turns out to be unstable nearly in
the whole range of $n$ where the difference
$\theta_2-\theta_1=\pi-2\theta_1$ with $\theta_1$ from
Eq.~\eqref{thetaMin} is non-zero. The situation remains
qualitatively the same, if in Eq.~\eqref{Etot2} for $E_{tot}$ we
take $\varepsilon_0(n)$ calculated using the density of
states~\eqref{rho0}. The behavior of $E_{tot}(n)$ in 2D case is
illustrated in Fig.~\ref{FigEtot2D}.

In three dimensions, the situation is more complicated. At small
doping, we have $\varepsilon_0(n)\approx-n+an^{5/3}$, where $$
a=\frac35\left(\frac{\pi^2}{\sqrt{6}}\right)^{2/3}\!\!\!, $$ and
the total energy becomes \begin{equation}\label{Etot3Da}
E_{tot}\approx-zt'n-z\left(\frac{t^2}{4J}-\frac{a}{n^{1/3}}
\right)n^2-\frac{zJ}{2}\,.
\end{equation} The second derivative of $E_{tot}$ is positive at
$n\to0$, but it changes sign at \begin{equation}
n_c\approx\left(\frac{5aJt'}{9t^2}\right)^3. \end{equation} Taking
into account the same arguments as in 2D case, we get an estimate
for the phase-separation range \begin{equation}\label{une2}
n_c\lesssim n\lesssim\frac{2J}{t}. \end{equation} We see that the
presence of nonzero nondiagonal hopping $t'$ leads to the
appearance of a lower critical concentration $n_c$ for phase
separation. (Maxwell construction would lead to phase separation
is a somewhat broader doping range, starting from some $n_0<n_c$).

Note that inequalities~\eqref{une1} and~\eqref{une2} are valid at
relatively small values of $J/t$ ratio.

\section{Anisotropic model}

Now we study a more realistic model of $e_g$ orbitals on the
square 2D lattice. This situation is characteristic, for example,
for layered cuprates, like K$_2$CuF$_4$, or manganites
(La$_2$MnO$_4$ or La$_2$Mn$_2$O$_7$). We assume that  an orbital
exchange Hamiltonian has Heisenberg-like form~\eqref{H_Heis}. In
the case of $e_g$ orbitals, any orbital can be written as a linear
combination of two basis functions $|a>=|x^2 - y^2>$ and
$|b>=|2z^2 - x^2 - y^2>$:
$|\theta>=\cos(\theta/2)|x^2-y^2>+\sin(\theta/2)|2z^2-x^2-y^2>$.
The hopping integrals $t^{\alpha\beta}$ in Eq.~\eqref{H_el} now
depend on the direction of hopping, and can be written in the form
of a matrix \begin{eqnarray}\label{t_xy}
(t_{x,y})^{\alpha\beta} =\frac{t_0}{4}\left(%
\begin{array}{cc}
3 & \mp\sqrt{3} \\
\mp\sqrt{3} & 1 \\
\end{array}%
\right)\,,
 \end{eqnarray}
where minus (plus) sign corresponds to $x$ ($y$) direction of
hopping.

Assuming again an underlying orbital structure corresponding to the
alternation of $|\theta_1\rangle$ and $|\theta_2\rangle$ orbitals,
we obtain the spectrum of charge carriers in the form
\begin{equation}\label{specEg}
E(\mathbf{k})=-t_0\left(A_{x}(\theta_1,\theta_2)\cos
k_x+A_{y}(\theta_1,\theta_2)\cos k_y\right)\,,
\end{equation}
where \begin{equation}\label{Axy}
A_{x,y}(\theta_1,\theta_2)=\left|\cos\left(\frac{\theta_1-\theta_2}
{2}\right)+\cos\left(\frac{\theta_1+\theta_2}{2}
\pm\frac{\pi}{3}\right)\right|.
\end{equation}
The total energy then reads
\begin{eqnarray}\label{EtotAn}
E_{tot}(\theta_1,\theta_2)&=&t_0\left(A_{x}(\theta_1,\theta_2)
+A_{y}(\theta_1,\theta_2)\right)\times\nonumber\\%
&&\varepsilon(n;\theta_1,\theta_2)+2J\cos(\theta_1-\theta_2)\,,
\end{eqnarray}
where $\varepsilon(n;\theta_1,\theta_2)=
\int_{-1}^{\mu}dE\,E\rho(n;\theta_1,\theta_2)$,
and the density of states can be written as
\begin{eqnarray}\label{rho12}
\rho(n;\theta_1,\theta_2)&=&\!\int\limits_0^
{\infty}\!\frac{ds}{\pi}\cos(Es)\times\\%
&&J_0\!\left(\frac{sA_x}{A_x+A_y}\right)
J_0\!\left(\frac{sA_y}{A_x+A_y}\right)\!.\nonumber \end{eqnarray}
Note that the density of states now depends on angles $\theta_1$,
$\theta_2$ via functions $A_{x,y}(\theta_1,\theta_2)$. In order to
find orbital structure, one should minimize $E_{tot}$,
Eq.~\eqref{EtotAn}, with respect to $\theta_1$ and $\theta_2$. The
analysis shows, that at doping $n$ less than some critical value
$n_1$, depending on the ratio $J/t_0$, the minimum of the total
energy corresponds to $\theta_{1}=0$, $\theta_{2}=\pi$, that is,
we have the homogeneous antiferro-orbital structure with
alternating $|x^2-y^2>$ and $|2z^2-x^2-y^2>$ orbitals (we ignore
here anharmonic effects and higher-order interactions, which
usually stabilize locally elongated octahedra with the angles, in
our notation, $\theta=\pi$, $\pm 2\pi/3$, see
Refs.~\onlinecite{kanamori,DK}). The energy of such a state is
\begin{equation} E_{tot}=t_0\sqrt{3}\varepsilon_0(n)-2J\,.
\end{equation} This state is locally stable,
$\partial^2E_{tot}/\partial n^2>0$.

\begin{figure} \begin{center}
\includegraphics*[width=0.45\textwidth]{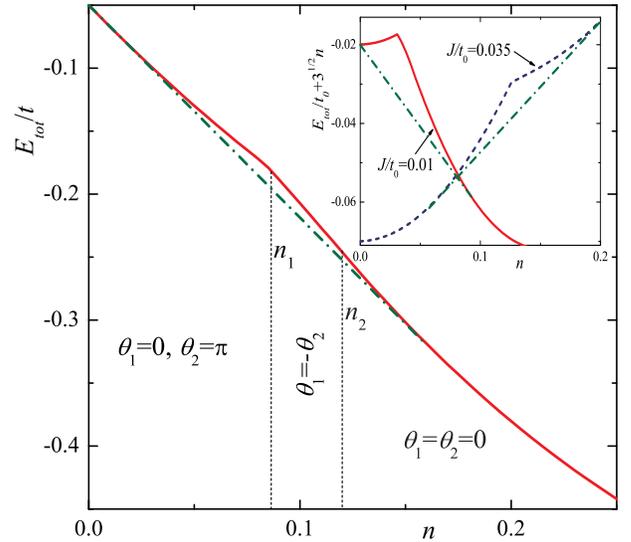} \end{center}
\caption{\label{FigEtot2DEg} (Color online) The energy of
homogeneous state for the anisotropic model as a functions of
doping at $J/t_0=0.025$ (red solid curve). In the region near
$n_1\approx0.08$, the homogeneous state is unstable toward a phase
separation. In the inset, the dependence of
$E_{tot}(n)+t_0\sqrt{3}n$ (linear term of the dependence
$E_{tot}(n)$ in the range $n<n_1$ is subtracted) on doping $n$ is
shown at the range near $n_1$ at different model parameters. The
red solid curve (blue dashed curve) corresponds to $J/t_0=0.01$
($J/t_0=0.035$), and have a negative (positive) value of
$\partial^2E_{tot}/\partial n^2$ in the region $n>n_1$ close to
$n_1$. The phase separation exists for both situations. Maxwell
construction is shown by dot-dashed line.} \end{figure}

At $n=n_1$, a jump-like transition to the canted state with
$\theta_2=-\theta_1$ occurs, where
\begin{equation}
\theta_1=\arccos\left(\frac{t_0|\varepsilon_0(n)|}{4J}\right)\,,
\end{equation} and $E_{tot}(n)$ has a kink at $n=n_1$. The energy
of such canted state at $n>n_1$ is
\begin{equation}\label{EtotEg}
E_{tot}=t_0\varepsilon_0(n)-\frac{t_0^2}{4J}\varepsilon_0^2(n)-2J\,.
\end{equation} With the further growth of $n$, the angle
$\theta_1$ decreases, and at $n=n_2$, determined by the equation
$t_0|\varepsilon_0(n_2)|/4J=1$, it vanishes, $\theta_1=0$
(ferro-OO with $|x^2-y^2\rangle$ orbitals). The total energy of
the system as function of doping is shown in
Fig.~\ref{FigEtot2DEg}. Note, that depending on the values of
parameters, the energy~\eqref{EtotEg} can have either positive or
negative curvature (see the inset to Fig.~\ref{FigEtot2DEg}). In
the former case, the homogeneous state is locally stable in the
whole range of doping, but the phase separation still exists in
the range of $n$ near $n=n_1$, due to the kink in the system
energy. In the second case, PS, of course, also exists (we have an
instability in some range of doping, where
$\partial^2E_{tot}/\partial n^2<0$). Note, that these two possible
situations (negative curvature of $E_{tot}$ and the kink) can lead
to inhomogeneous states with quite different
properties~\cite{DiCastro}.

\section{Inhomogeneities in the orbitally ordered structures}\label{Inhom}

We demonstrated above that the additional charge carriers
introduced  to the orbitally ordered structures can lead to the
formation of an inhomogeneous state. Now, let us discuss possible
types of such inhomogeneities in more detail using a model of the
$e_g$ orbitals at the sites of 2D square lattice, considered in
the previous Section. We assume that each charge carrier forms a
finite region of an OO structure with alternating
$|\theta_1\rangle$ and $|\theta_2\rangle$ orbitals (not
necessarily ferro-OO with $\theta_1=\theta_2=0$) to optimize
$H_{\text{el}}$. The remaining part of the crystal has
antiferro-OO structure with $|x^2 - y^2>$ and $|2z^2 - x^2 - y^2>$
orbitals, according to the results of the previous Section at
$n\to0$.

\begin{figure}[htp]\centering
   \subfigure[$J/t_0 = 0.005$]{
      \includegraphics[width=0.45\textwidth]{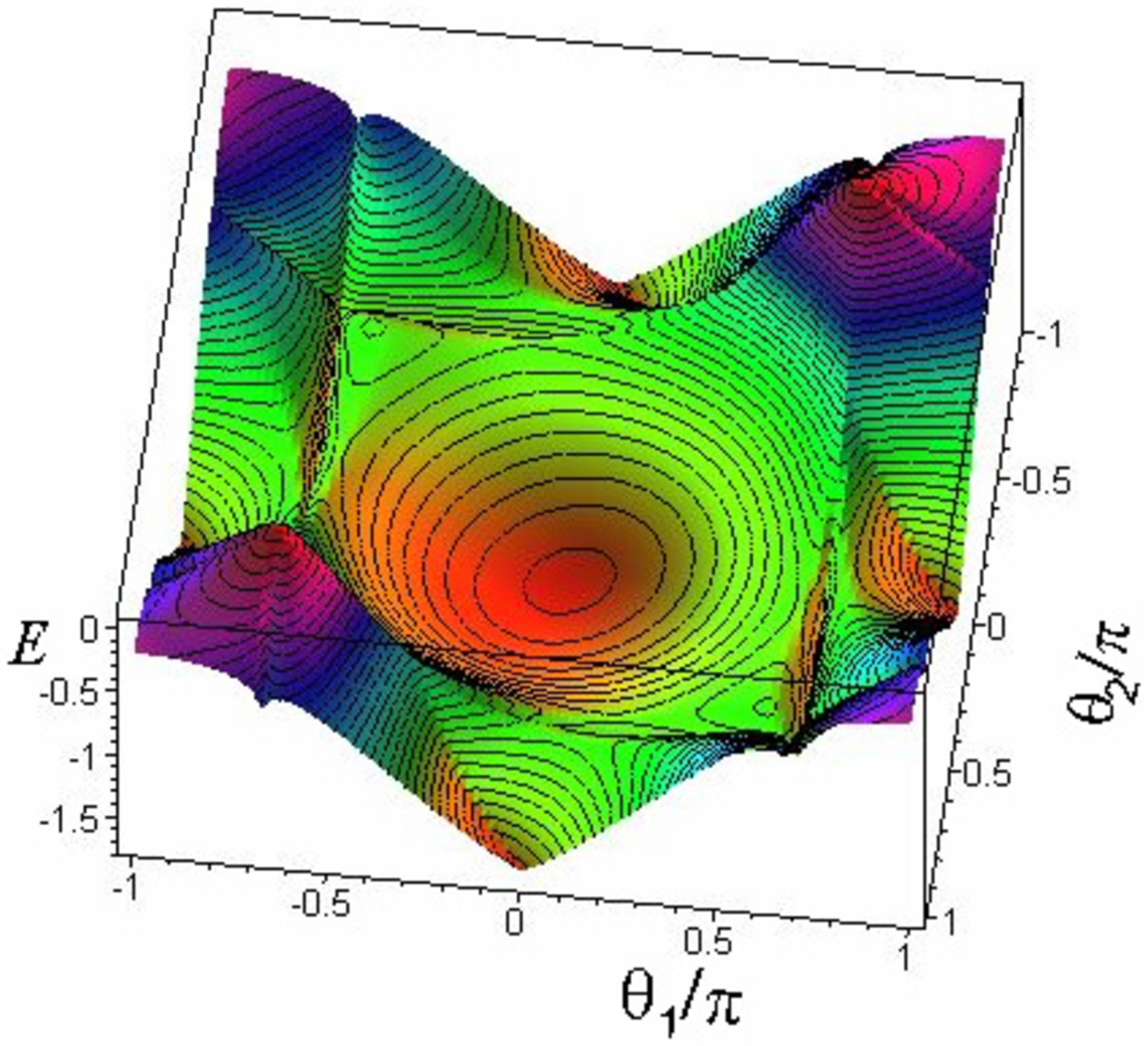}}\\
   \subfigure[$J/t_0 = 0.02$]{
      \includegraphics[width=0.45\textwidth]{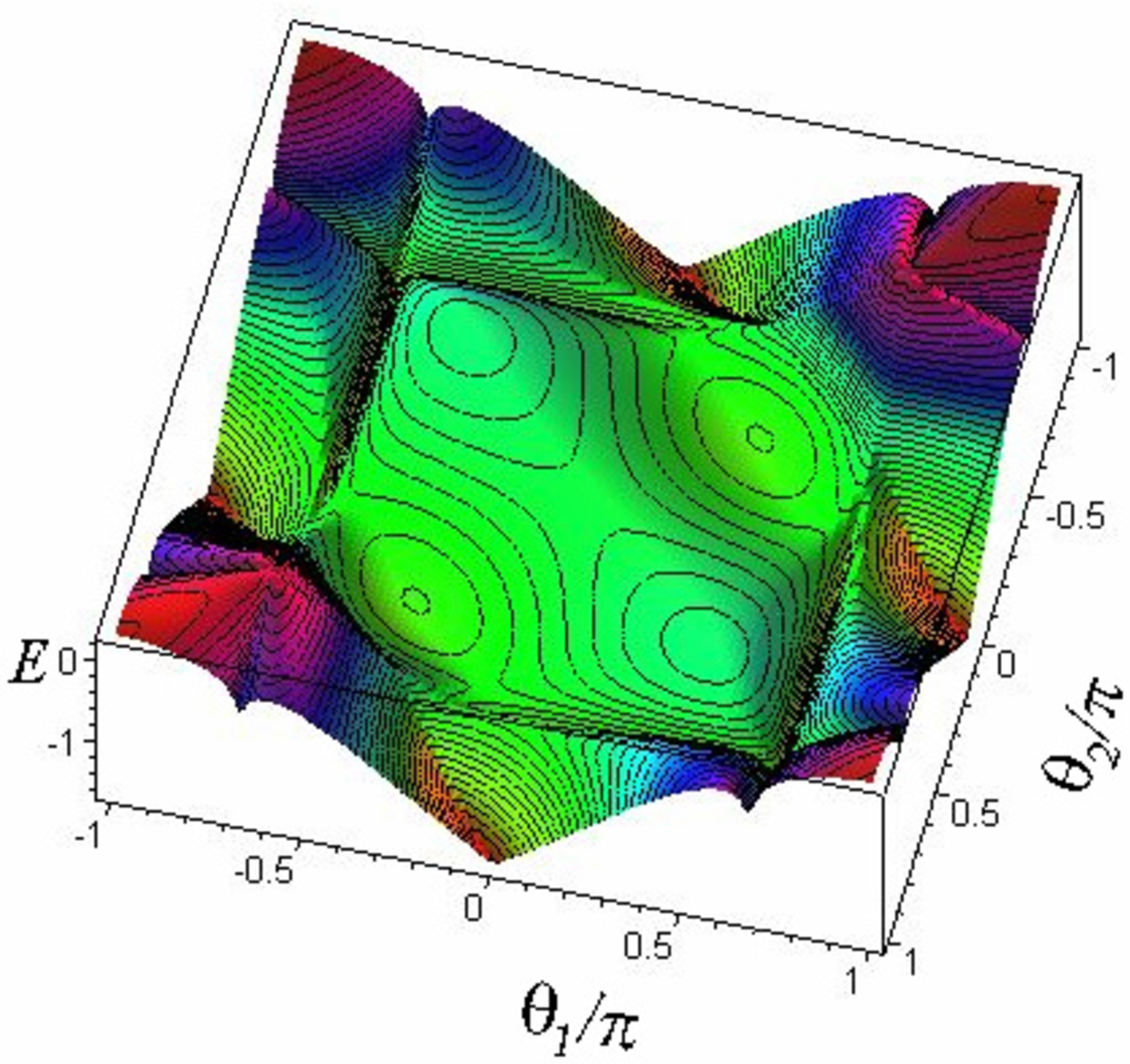}}
\caption{Total energy as function of angles $\theta_1$ and
$\theta_2$ at $J/t_0$ smaller (a) and larger (b) than the critical
value $\simeq 0.0075$.} \label{FigE3D}
\end{figure}

The spectrum of charge carriers is given by Eq.~\eqref{specEg}.
Expanding this spectrum in power series of $\mathbf{k}$ up to the
second order, we find an effective Hamiltonian for a charge
carrier in a finite region:
\begin{equation}\label{Heff}
\hat{H}_{eff} = -t_0\left(A_x + A_y\right)%
+\frac{t_0}{2}\left(A_x\frac{\partial^2}{\partial x^2}
+A_y\frac{\partial^2}{\partial y^2}\right), \end{equation} where
$A_x$, $A_y$ are given by Eq.~\eqref{Axy}. Using
Hamiltonian~\eqref{Heff}, we can solve the Schr\"{o}dinger
equation within a finite region, which we choose in the shape of
ellipse with semiaxes $\sqrt{A_x}\rho_0$ and $\sqrt{A_y}\rho_0$.
As a result, we find the following expression for the kinetic
energy of the charge carrier within such droplet
\begin{equation}\label{Ekin} E_{kin} = -t_0\left(A_x +
A_y\right)+\frac{t_0j_{0,1}^2}{2\rho_0^2}\,, \end{equation} where
$j_{0,1}\cong2.405$ is the first root of Bessel function $J_0$.
The potential energy $E_{pot}$ related to the orbital ordering is
the sum of two contributions proportional to the droplet volume
$v$ ($v=\pi\sqrt{A_xA_y}\rho_0^2$): the energy of the canted OO
within the droplet is $zJ\cos(\theta_1-\theta_2)/2$ and the loss
in energy of the antiferro-OO matrix due to the formation of the
droplet is $zJ/2$. As a result, we get ($z=4$)
\begin{equation}\label{Epot}
E_{pot}=4\pi\rho_0^2J\sqrt{A_xA_y}\cos^2
\left(\frac{\theta_1-\theta_2}{2}\right)\,. \end{equation}
Minimizing the droplet energy $E_{kin}+ E_{pot}$ with respect to
$\rho_0$, we find \begin{equation}\label{rho_0} \rho_0
=\left(\frac{t_0j_{0,1}^2}{8\pi J\sqrt{A_xA_y}
\cos^2\left(\frac{\theta_1-\theta_2}{2}\right)}\right)^{1/4}\!\!.
\end{equation} The total energy (per lattice site) then reads
\begin{eqnarray}\label{Etot(theta)}
E_{tot}&=&-2J+E(\theta_1,\,\theta_2)\,n\,,\\
E(\theta_1,\,\theta_2)&=&-2t_0(A_x+A_y)\phantom{\frac12}\\
&&+j_{0,1}\left(8\pi
t_0J\sqrt{A_xA_y}\right)^{1/2}\left|\cos
\left(\frac{\theta_1-\theta_2}{2}\right)\right|,\nonumber
\end{eqnarray}
where we assume that all charge carriers introduced by doping form
such identical OO droplets.

\begin{figure*} \begin{center}
\includegraphics*[width=0.48\textwidth]{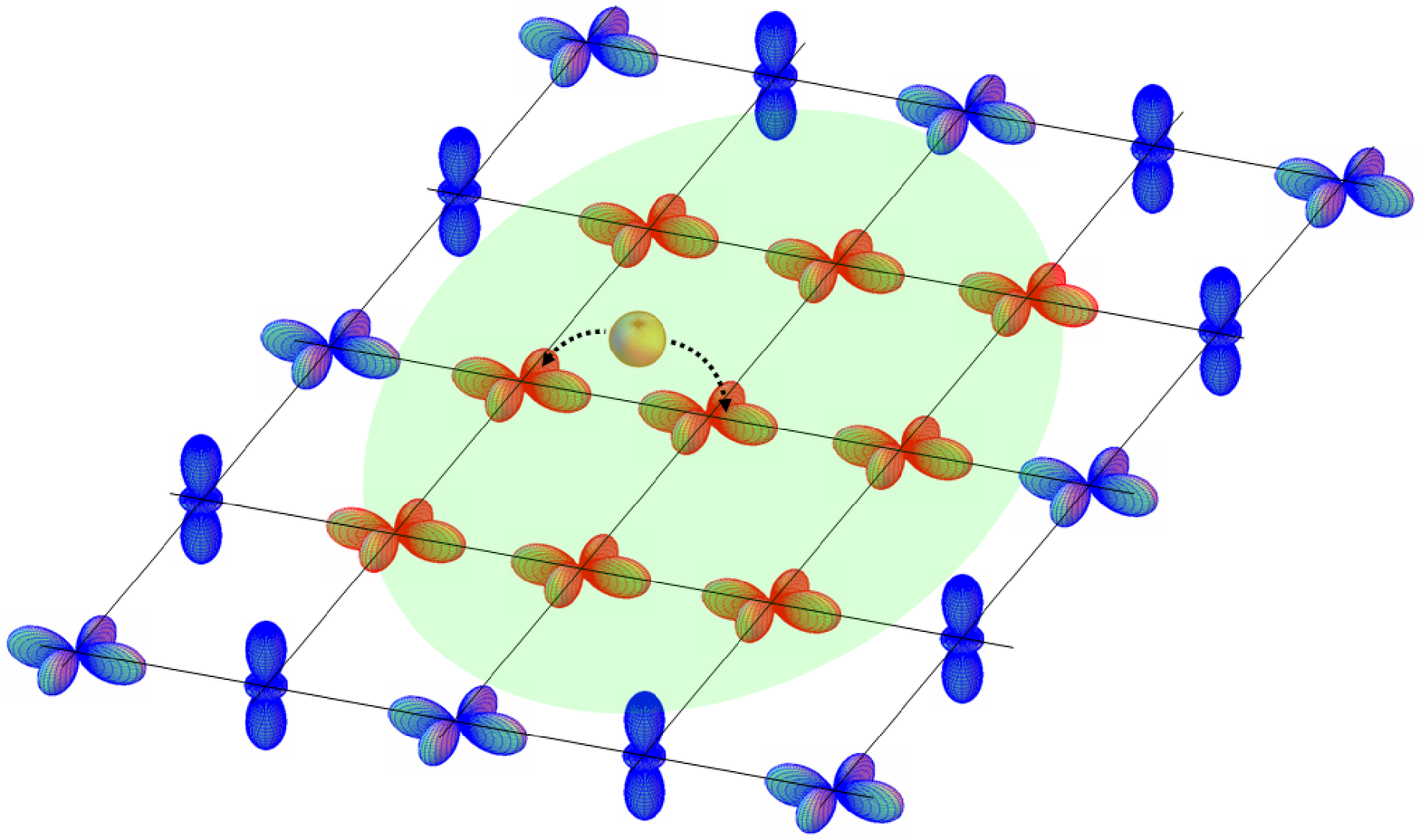}
\includegraphics*[width=0.48\textwidth]{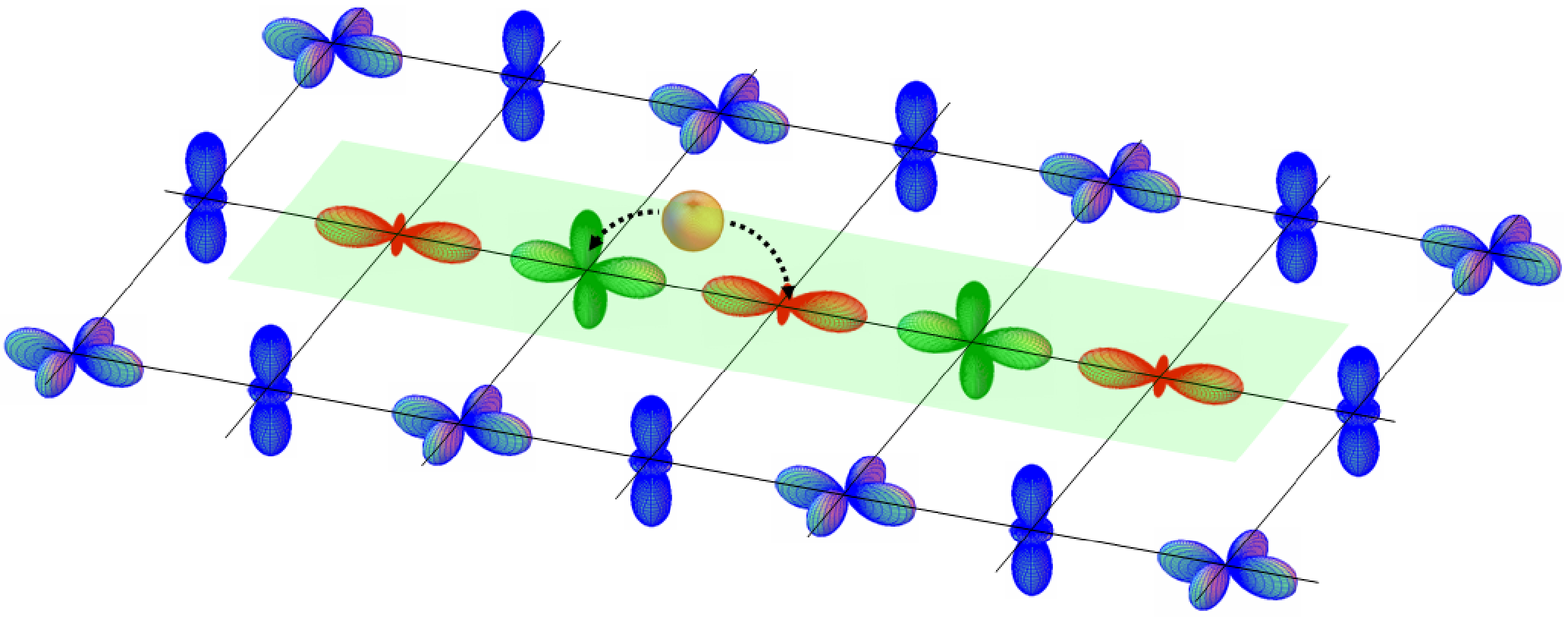}
\end{center} \caption{\label{FigPolaron} (Color online) Schematic
illustration of circular (left panel) and needle-like (right
panel) droplets. An electron or hole moves in a finite region
creating a ferro- or canted-OO structure, in antiferro-OO matrix.
In the case of hole, there exists a one mobile empty site within a
droplet.} \end{figure*}

To find possible types of OO droplets, we minimize $E$ with
respect to $\theta_1$ and $\theta_2$ (note again, that functions
$A_{x,y}$ depend on $\theta_1$, $\theta_2$ according to
Eq.~\eqref{Axy}). The function $E(\theta_1,\,\theta_2)$ at two
different values of $J/t_0$ is shown in Fig~\ref{FigE3D}. In
general case, the function $E(\theta_1,\,\theta_2)$ has several
minima, and the values of $\theta_1$ and $\theta_2$ corresponding
to the lowest minimum depend drastically on parameter $J/t_0$. At
small $J/t_0$ (Fig~\ref{FigE3D}a), the lowest minimum corresponds
to $\theta_1=\theta_2=0$, that is, we have ferro-OO structure
inside the droplet with occupied $|x^2-y^2>$ orbitals. In this
case, the most favorable shape of droplets is a circle (see left
panel of Fig.~\ref{FigPolaron}). At $J/t_0$ larger than some
critical value ($J_{cr}/t_0\simeq0.0075$), the minimum
$\theta_1=\theta_2=0$ becomes metastable, and the energy
$E(\theta_1,\,\theta_2)$ has four degenerate lowest minima: two of
them correspond to $\theta_1=\pi/3$, $\theta_2=2\pi/3$,
$\theta_1=-\pi/3$, $\theta_2=-2\pi/3$, and the similar two minima
with the replacement $\theta_1\leftrightarrow\theta_2$ (see
Fig~\ref{FigE3D}b). In this case, we have chains of alternating
$|x^2-z^2>$ and $|2y^2 -x^2-z^2>$ (or $|y^2-z^2>$ and $|2x^2
-y^2-z^2>$) orbitals and hence nearly one-dimensional
(cigar-shaped) droplets stretched along $y$ or $x$ axes (right
panel of Fig.~\ref{FigPolaron}). With the further growth of
$J/t_0$ the metastable state $\theta_1=\theta_2=0$ splits into two
states corresponding to $\theta^*_1=-\theta^*_2$ with positive and
negative $\theta^*_1$, as it can be seen from Fig~\ref{FigE3D}b.
The droplets corresponding to these states have circular shape,
but canted OO structure.

The existence of two types of droplets with different shapes can
be easily understood. The maximum gain in the kinetic energy
corresponds to the ferro-OO state with $|x^2-y^2>$ orbitals. At
small $J/t_0$, the kinetic energy prevails, and we have circular
droplets with this type of orbitals. The minimum cost in the
potential energy corresponds to nearly one-dimensional structures.
At larger $J/t_0$, the potential energy plays more important role
than the kinetic one, and we get cigar-like droplets (smaller
volume of such a droplet gives smaller loss of orbital energy).
The orbital structure inside the droplet described above
corresponds to the maximum gain in the kinetic energy for
one-dimensional chain (in the absence of hopping between
neighboring chains).

The analysis shows, that the energy of an inhomogeneous state,
Eq.~\eqref{Etot(theta)}, consisting of circular or cigar-like OO
droplets embedded into an antiferro-OO matrix is less than the
energy of a homogeneous state in a certain range of doping
$0<n<n^*_c$. With the growth of the number of charge carriers, the
droplets start to overlap, and at $n=n^*_c$ the inhomogeneous
state of considered type (ferro-OO droplets in antiferro-OO
matrix) disappears. However, the phase separation exists in a
wider range of doping (see the previous Section). For circular
droplets, we have an estimate $n^*_c\sim1/\pi\rho_0^2$. Taking for
estimate the ratio $J/t_0=0.005$, we get $\rho_0\approx2$ (in
units of lattice constant) and $n^*_c\sim0.08$.

In the case of cigar-like droplets, we have $A_x=0$ (or $A_y=0$),
and according to Eq.~\eqref{rho_0} we would get that chains have
infinite length (but zero volume $v$), $\rho_0=\infty$.  This is
of course not a very realistic result, coming from an
approximation, where the potential is assumed to be proportional
to the droplet volume only. In order to estimate the
characteristic length $L$ of the chain, we should take into
account the surface term (proportional to the droplet's length) in
the potential energy $E_{pot}$ of the droplet. Let us consider,
for definiteness, the chain of $|y^2-z^2>$ and $|2x^2 -y^2-z^2>$
orbitals, stretched along $x$ axis. In this case, we have
$A_x=\sqrt{3}$, $A_y=0$. The effective Hamiltonian \eqref{Heff} is
reduced to $$ \hat{H}_{eff}=-t_0\sqrt{3}+\frac{t_0\sqrt{3}}{2}
\frac{\partial^2}{\partial x^2}\,, $$ and the kinetic energy of
the charge carrier in the chain of length $L$  becomes
$E_{kin}=-t_0\sqrt{3}(1-\pi^2/2L^2)$. The surface energy of
interorbital exchange interaction has a minimum, when the chain is
located in an antiferro-OO matrix like shown in
Fig.~\ref{FigPolaron}: each $|y^2-z^2>$ ($|2x^2 -y^2-z^2>$)
orbital in the chain has its nearest neighbor $|x^2-y^2>$ ($|2z^2
-x^2-y^2>$) orbital in the matrix. In continuum approximation, the
potential energy can be written as $E_{pot}=9JL/4$. Minimizing
$E_{kin}+E_{pot}$ with respect to $L$, we arrive at the following
formula for characteristic length of the chain:
\begin{equation}\label{L0} L_0
=\left(\frac{4\pi^2t_0\sqrt{3}}{9J}\right)^{1/3}\!\!.
\end{equation} At $J/t_0=0.05>J_{cr}/t_0$, we have
$L_0\approx5.5$. At random distribution  of the chains in the
matrix (we have chains stretched both along $x$ and $y$ axes), the
critical concentration is about $n^*_c\sim1/L_0^2$, but it can be
larger if a more complicated structure of chains, e.g. regular
stripes, appears in the system.

\section{Conclusions}

We have studied a simple model of electronic phase separation in
the system of charge carriers moving in an orbitally ordered
background. It was shown that a homogeneous state in such a system
can be unstable toward a phase separation, where delocalized
charge carriers favor the formation of nanoscale inhomogeneities
with the orbital structure different from that in the undoped
material. The shapes and sizes of such inhomogeneities were
determined for 2D lattice of $e_g$ orbitals. The shape of
inhomogeneities depends drastically on the ratio of interorbital
exchange interaction and a hopping amplitude of the charge
carriers, $J/t_0$: there exists a critical value of $J/t_0$,
corresponding to the transition from the circular inhomogeneities
to a one-dimensional chains of finite length.

The model under study is quite similar to the double exchange
model,  where the orbital variables play a role of local spins. It
is well known that such a model also exhibits an instability
toward a phase separation into phases with different types of
magnetic ordering. The inhomogeneous state with circular ferro-OO
droplets is, in essence, an analog of a magnetic polaron state
(ferromagnetic droplets in an antiferromagnetic matrix), which is
usually considered in the double exchange
model~\cite{Nag,Kak,KaKhoMo}. Nevertheless, our orbital model is
more complicated than the usual double exchange due to the
existence of non-diagonal hopping amplitudes and to the anisotropy
in hoppings. Both these features lead to the results specific for
the orbital model, such as the kink in the energy of a homogeneous
state and canted-OO needle-like droplets.

In the present paper, any magnetic structure and spins of the
charge carriers were fully neglected. Taking into account spin
degrees of freedom can lead to the formation of inhomogeneities
with different orbital and spin configurations.

In the proposed model, the localized electrons forming an orbital
order  and the conduction electrons or holes were supposed to be
two different groups of electrons. However, we can argue that our
main results are also valid for a model, where the same electrons
take part both in the hopping and in the formation of orbital
ordered structure. Indeed, in the case of magnetic oxides with
Jahn-Teller ions, an orbital degeneracy is lifted by lattice
distortions, giving rise to an orbitally-ordered ground state at
$n=1$. If we suppose that a long-range orbital ordering still
exists at small hole doping $x=1-n\ll 1$, we come to the situation
considered in present paper: we have holes moving in an
orbitally-ordered background. In a mean-field approximation, we
only should replace in all formulas above $n\to x=1-n$ and $J\to
J(1-x)^2$, since the number of sites taking part in interorbital
exchange interaction is reduced by a factor of $1-x$ . In the
materials with Jahn-Teller ions, the orbital Hamiltonians are more
complicated than the Heisenberg-like Hamiltonian considered in
this paper. Preliminary calculations for the Hamiltonian
corresponding to the superexchange mechanism of the orbital
ordering~\cite{KK} show that the obtained results remain
qualitatively the same. However, in real substances, there also
exists a possibility of OO without local distortions,
corresponding to complex combinations of $e_g$
orbitals~\cite{vBrK2001}, which needs a special analysis.

\section*{Acknowledgments}

The work was supported by the European project CoMePhS (contract
NNP4-CT-2005-517039), International Science and Technology Center
(grant G1335), Russian Foundation for Basic Research (projects
07-02-91567 and 08-02-00212), and by the Deutsche
Forschungsgemeinshaft via SFB 608 and the German-Russian project
436 RUS 113/942/0. A.\,O. also acknowledges support from the
Russian Science Support Foundation.


\begin{thebibliography} {99}

\bibitem[\S]{affUK}
Also at the Department of Physics, Loughborough University,
Leicestershire, LE11 3TU, UK.

\bibitem{KK}
K.I. Kugel and D.I. Khomskii, Usp. Fiz. Nauk {\bf 136}, 621 (1982) [Sov.
Phys.—Uspekhi {\bf 25}, 231 (1982)]

\bibitem{KaVe} M.D. Kaplan and B.G. Vekhter,  \textit{Cooperative
Phenomena in Jahn-Teller Crystals} (Plenum, New York, 1995).

\bibitem{dagbook} E. Dagotto, {\em Nanoscale Phase Separation and
Colossal Magnetoresistance: The Physics of Manganites and Related
Compounds} (Springer-Verlag, Berlin, 2003).

\bibitem{Nag} E. Nagaev, \textit{Colossal Magnetoresistance and
Phase Separation in Magnetic Semiconductors} (Imperial College
Press, London, 2002).

\bibitem{Kak}
M.Yu. Kagan and K.I. Kugel, Usp. Fiz. Nauk. {\bf
171}, 577 (2001) [Physics - Uspekhi {\bf 44}, 553 (2001)].

\bibitem{KilKha} R. Kilian and G. Khaliullin, \prb {\bf 60}, 13458
(1999).

\bibitem{MiKhoSaw} T. Mizokawa, D.I. Khomskii, and G.A. Sawatzky,
\prb {\bf 63}, 024403 (2000).

\bibitem{KhaOk} G. Khaliullin and S. Okamoto, \prl {\bf 89},
167201 (2002).

\bibitem{vdBKhaKho} J. van den Brink, G. Khaliullin, and D.
Khomskii, Orbital effects in manganites, Ch. 6 in \textit{Colossal
Magnetoresistive Manganites}, ed. T. Chatterji, (Kluwer,
Dordrecht, The Netherlands, 2004), pp. 263-302.

\bibitem{KaKhoMo} M.Yu. Kagan, D.I. Khomskii, and M.V. Mostovoy,
Eur. Phys. J. B {\bf 12}, 217 (1999).

\bibitem{kanamori} J. Kanamori, J. Appl. Phys. {\bf 31}, 14S
(1960).

\bibitem{DK} D. Khomskii and J. van den Brink, \prl {\bf 85}, 3329
(2000).

\bibitem{DiCastro} C. Ortix, J. Lorenzana, and C.~Di~Castro,
arXiv:0801.0955 (2008).

\bibitem{vBrK2001} J. van den Brink and D. Khomskii, \prb {\bf
63}, 140416(R) (2001).

\end{thebibliography}
\end{document}